\documentclass[a4paper]{IEEEtran}
\IEEEoverridecommandlockouts
\usepackage[utf8]{inputenc}
\usepackage{subcaption}
\usepackage{amsmath,amssymb,amsfonts, bm}

\usepackage[linesnumbered,lined,boxed,commentsnumbered, ruled]{algorithm2e}
\usepackage{graphicx}
\usepackage{makecell}
\usepackage{multirow}
\usepackage{cite}
\usepackage{fouriernc}
\DeclareMathAlphabet{\mathcal}{OMS}{zplm}{m}{n}
\usepackage[table,xcdraw]{xcolor}
\usepackage{textcomp}
\usepackage{lipsum}
\usepackage{comment}
\usepackage{csquotes}
\usepackage{array}
\usepackage{xcolor}
\usepackage{url}
\usepackage{diagbox}
\usepackage[nameinlink,capitalise]{cleveref}
\usepackage{enumitem}
\def\BibTeX{{\rm B\kern-.05em{\sc i\kern-.025em b}\kern-.08em
    T\kern-.1667em\lower.7ex\hbox{E}\kern-.125emX}}

\usepackage[acronym,toc,nonumberlist]{glossaries}
\makeglossaries
\newacronym{5G}{5G}{Fifth Generation}
\newacronym{MIMO}{MIMO}{Multiple Input - Multiple Output}
\newacronym{SISO}{SISO}{single-input-single-output}
\newacronym{C-RAN}{C-RAN}{Centralized Radio Access Network}
\newacronym{TDM}{TDM}{time division multiplexing}
\newacronym{CoMP}{CoMP}{Coordinated Multi-Point}
\newacronym{RRU}{RRU}{remote radio unit}
\newacronym{BBU}{BBU}{baseband unit}
\newacronym{OPEX}{OPEX}{operating cost}
\newacronym{CAPEX}{CAPEX}{Capital cost}
\newacronym{CPRI}{CPRI}{Common Public Radio Interface}
\newacronym{eCPRI}{eCPRI}{eCommon Public Radio Interface}
\newacronym{OBSAI}{OBSAI}{Open base station architecture initiative}
\newacronym{Gbps}{Gbps}{Giga bits per second}
\newacronym{UE}{UE}{user equipmenFiber}
\newacronym{LTE}{LTE}{Long Term Evolution}
\newacronym{MMSE}{MMSE}{Minimum Mean Squared Error}
\newacronym{ZF}{ZF}{Zero Forcing}
\newacronym{LS}{LS}{Least Square}
\newacronym{BER}{BER}{bit error rate}
\newacronym{SNR}{SNR}{signal to noise ratio}
\newacronym{SINR}{SINR}{signal to interference and noise ratio}
\newacronym{PLC}{PLC}{Power-Line Communication}
\newacronym{AWGN}{AWGN}{additive white Gaussian noise}
\newacronym{PSD}{PSD}{power spectral density}
\newacronym{HARQ}{HARQ}{Hybrid automatic repeat request}
\newacronym{ARQ}{ARQ}{automatic repeat request}
\newacronym{NR}{NR}{New Radio}
\newacronym{PDU}{PDU}{power distribution unit}
\newacronym{FEC}{FEC}{forward error correction}
\newacronym{BW}{BW}{bandwidth}
\newacronym{CU}{CU}{Central Unit}
\newacronym{DU}{DU}{Distributed Unit}
\newacronym{RU}{RU}{Radio Unit}
\newacronym{AU}{AU}{Antenna Unit}
\newacronym{CSMA_CD}{CSMA/CD}{Carrier Sense Multiple Access/ Collision Detection}
\newacronym{SVD}{SVD}{singular value decomposition}
\newacronym{SR}{SR}{Selective repeat}
\newacronym{RS-FEC}{RS-FEC}{Reed-Solomon Forward Error Correction}
\newacronym{CRC}{CRC}{cyclic redundancy check}
\newacronym{C_M}{C\&M}{Control and Management}
\newacronym{NRZ}{NRZ}{non-return to zero}
\newacronym{PAPR}{PAPR}{peak to average power ratio}
\newacronym{FER}{FER}{frame error rate}
\newacronym{IND-Re}{IND-Re}{Impulsive noise detection/re-transmission}
\newacronym{uavs}{UAVs}{Unmanned Aerial Vehicles}
\newacronym{WSN}{WSN}{wireless sensor network}
\newacronym{MGDC}{MGDC}{Minimum Geometric Disk Cover}
\newacronym{TSP}{TSP}{Traveling Salesman Problem}
\newacronym{haps}{HAPS}{High Altitude Platform Station}
\newacronym{LoS}{LoS}{Line of Sight}
\newacronym{ILP}{ILP}{integer linear programming}
\newacronym{MILP}{MILP}{mixed integer linear programming}
\newacronym{INCM}{INCM}{IoT Node Clusters Minimization}
\newacronym{CM}{CM}{Cluster Minimization}
\newacronym{SCP}{SCP}{Set Covering Problem}
\newacronym{SCOP}{SCOP}{Set Covering Optimization Problem}
\newacronym{leach}{LEACH}{Low-Energy Adaptive Clustering Hierarchy}
\newacronym{IoT}{IoT}{Internet of Things}
\newacronym{SC}{SC}{Segment Clustering}
\newacronym{CETSP}{CETSP}{Close Enough Traveling Salesman Problem}
\begin{document}

\title{On Achieving High-Fidelity Grant-free Non-Orthogonal Multiple Access}

\author{\IEEEauthorblockN{ 
Haoran Mei,
Limei Peng,
and Pin-Han Ho~\IEEEmembership{IEEE Fellow}
}\\
\thanks{Haoran Mei and Limei Peng are with the School of Computer Science and Engineering, Kyungpook National University, Deagu, South Korea (e-mail: \{meihaoran, auroraplm\}@knu.ac.kr).}
\thanks{Pin-Han Ho is with Shenzhen Institute for Advanced Study, UESTC, and Department of Electrical and Computer Engineering, University of Waterloo, Waterloo, ON, Canada (e-mail: p4ho@uwaterloo.ca).}


}

\maketitle
\begin{abstract} 
Grant-free access (GFA) has been envisioned to play an active role in massive Machine Type Communication (mMTC) under 5G and Beyond mobile systems, which targets at achieving significant reduction of signaling overhead and access latency in the presence of sporadic traffic and small-size data. The paper focuses on a novel K-repetition GFA (K-GFA) scheme by incorporating Reed-Solomon (RS) code with the contention resolution diversity slotted ALOHA (CRDSA), aiming to achieve high-reliability and low-latency access in the presence of massive uncoordinated MTC devices (MTCDs). We firstly defines a MAC layer transmission structure at each MTCD for supporting message-level RS coding on a data message of $Q$ packets, where a RS code of $KQ$ packets is generated and sent in a super time frame (STF) that is composed of $Q$ time frames. The access point (AP) can recover the original $Q$ packets of the data message if at least $Q$ out of the $KQ$ packets of the RS code are successfully received. The AP buffers the received MTCD signals of each resource block (RB) within an STF and exercises the CRDSA based multi-user detection (MUD) by exploring signal-level inter-RB correlation via iterative interference cancellation (IIC). With the proposed CRDSA based K-GFA scheme, we provide the complexity analysis, and derive a closed-form analytical model on the access probability for each MTCD as well as its simplified approximate form. Extensive numerical experiments are conducted to validate its effectiveness on the proposed CRDSA based K-GFA scheme and gain deep understanding on its performance regarding various key operational parameters.

\end{abstract}
\begin{IEEEkeywords}
K-repetition Grant-free access (K-GFA), massive machine type communication (mMTC), Reed-Solomon (RS) code, interference cancellation (IC).
\end{IEEEkeywords}

\section{Introduction}
\IEEEPARstart{M}{assive} machine-type communication (mMTC), one of the three major services in 5G new radio (NR) as defined by the International Telecommunication Union (ITU), is designed to achieve massive connectivity while supporting high data rate and low-cost devices\cite{grant_free_NOMA_survey}. Under such a circumstance, the legacy grant-based access (GBA) approach may result in long delay and stringent limitations on the number of mMTC devices (MTCD) that can simultaneously access the network, leading to a substantial challenge in provisioning efficient and reliable uplink (UL) transmissions, particularly when dealing with a large number of MTCDs that communicate with short packets and sporadic traffic.

As a remedy, grant-free access (GFA) has attracted extensive attention from the research society as a graceful complement of the legacy GBA. With GFA, the MTCDs transmit their data without waiting for the grant from the access point (AP) \cite{2-rach, 4-rach}, thereby diminishing the access latency and signalling overhand to the best extent. Nonetheless, such saving is at the expense of possible collisions between two accessing MTCDs on a common resource block (RB), resulting in several issues on transmission reliability and the overall throughput/rate.

To mitigate the malicious effect of potential collisions in the GFA systems, K-repetition suggests to allow an MTCD to transmit a packet in $K$ replicas in each time frame. To explore the best repetition diversity and temporal diversity, the paper investigates a novel K-repetition GFA (K-GFA) scheme, in which the K-repetition mechanism is incorporated with contention resolution diversity slotted ALOHA (CRDSA) and Reed-Solomon (RS) code \cite{NC1, K-scma}, in order to achieve high-reliability and low-latency UL access in the presence of incongruous and uncoordinated resource selections of the MTCDs. Specifically, the proposed K-GFA scheme deploys a ($KQ$, $Q$) RS code on the data message and the codeword of a size $KQ$ is transmitted using a number of $KQ$ RBs in a super time frame (STF) that contains $Q$ time frames. With iterative interference cancellation (IIC), the AP buffers the received signals from all RBs and performs IC on each RB by taking the user signals already obtained in the previous iterations as the multiple access interference (MAI). 
Facilitated by the ($KQ$, $Q$) RS code, successful retrieval of the codeword is claimed if at least $Q$ out of the $KQ$ packets of the RS code within the STF are successfully obtained.

The contributions of the paper are given as follows:
\begin{itemize}
\item Introduce a novel K-GFA scheme that incorporates with a multi-user detection (MUD) mechanism based on CRDSA and RS code. 
\item Develop analytical models on access probability under the proposed K-GFA scheme. 
\item Conduct extensive numerical experiment to validate the proposed models, and gain deep understanding on the access probability and message delay performance of the proposed scheme by considering various key parameters.
\end{itemize}

The rest of the paper is organized as follows. Section II provides literature review. Section III presents the system model. Section IV provides detailed description of the proposed scheme. Section V introduces a generic implementation model of the proposed scheme under blind IC. Section VI presents our analytical models by allowing up to two iterations of IIC and one MTCD signal as for MAI. Section VII validates the proposed analytical model and gain deep understanding on the performance of the proposed scheme in terms of access probability and message delay. Section VIII concludes the paper.

\section{Literature Review}

\subsection{Contention Resolution (CR) for Random Access Channel}

Contention resolution (CR) in K-GFA can be achieved by employing a MUD scheme, and has been widely investigated in the following categories: power-based (PB), code-based (CB), compressed sensing (CS)-based, and machine learning (ML)-based \cite{grant_free_NOMA_survey}. 

The PB-based CR utilizes successive interference cancellation (SIC) to manipulate power-level differences of the signals at a RB for MUD. Combining ALOHA with PB-NOMA allows for adaptive power selection from a preset pool and thus the AP can better estimate the amount of active devices \cite{pd-mud2}. \cite{pd-mud3} presents an analytical model for establishing a lower bound on the system throughput where taking both the number of RBs and the available power levels are taken into consideration. However, PB-based CR is subject to decreased efficacy with the increasing number of active devices due to the hardness of finding viable power-level distributions among the contending MTCDs.

For CB-based CR, \cite{cd-based3} conducts a comprehensive study on codebook design and identifies that provisioning more codes can accommodate more devices at the expense of escalated receiver complexity. Similar observations and conclusions are reported in \cite{cd-based4}, where the MUD performance is notably influenced by some environmental factors such as noise and interference. Consequently, CB-based CR might not be suitable to the scenarios with numerous miniature MTCDs of limited capacity.

CS is explored for MUD due to sparse user activity in mMTC. It can be combined with message passing algorithms (MPA) for joint active user and data detection \cite{cs-based1}. \cite{cs-based3} introduced compressive sampling matching pursuit (CoSaMP) algorithm for sporadic transmissions. \cite{cs-based5} presents two detectors that integrate a generalized approximate message passing algorithm into sparse Bayesian learning (SBL) and pattern coupled sparse Bayesian learning (PCSBL) algorithms to achieve low-complexity CS-based CR. Although effective in some scenarios, the CS-based CR is subject to similar limitations as CB-based CR. 

Considering the increasing complexity with a growing number of devices, ML-based approaches are being explored for efficient MUD. \cite{ml-based1} employs cross-validation to determine user sparsity. \cite{ml-based2} utilizes deep learning to map the received signals to active users, achieving superior performance compared to conventional MUD algorithms. \cite{ml-based3} presents an attention-based bidirectional long short-term memory to achieve joint user and data detection, leveraging the device activation history and the complex spreading sequences. However, the ML-based approaches rely on labeled data for training, which may not always be available.

\subsection{Repetition Correlation-based MUD}
Repetition correlation serves as an alternative MUD method. In \cite{SIC1}, a signal processing module decodes collision-free signals and removes their replicas from the associated RBs. 
Although effective in some scenarios, it may introduce extra overhead and system complexity for accurately locating all signal replicas. As a remedy, CRDSA \cite{CRDSA1} has each MTCD to launch multiple replicas of each packet via randomly selected time slots during a random access time window. It assumes only ''clean" time slots, containing a single user's signal, are decodable. Enhancements on this class of schemes include power density-based SIC and its applications on non-orthogonal multiple access (NOMA) for heavy traffic scenarios \cite{CRDSA4}. Additionally, enhancing the CRDSA performance can also be achieved by implementing environment-aware adaptive control for the repetition strategies of MTCDs. One effective approach is to leverage deep reinforcement learning techniques \cite{DRL_K_adj}.

As a solid expansion of CRDSA, parallel interference cancellation (PIC) \cite{PIC1} has been proposed for achieving high access probability without requiring the power assignment among the user signals. With PIC, the received signal of each time slot in a random access time window is buffered at the AP first and its replica is decoded. The AP can parse the successfully decoded signal to obtain the information of which time slots else are taken for launching its signal copies, and the successfully decoded signal is taken as "interference" and removed from the corresponding buffered signals.

\section{System model}
For GFA, a two-step RACH procedure \cite{2-rach} is defined by 3GPP. This procedure involves message A (MsgA) carrying preamble and payload signals in the UL, while message B (MsgB) handling random access response (RAR) and contention resolution in the downlink (DL). An MTCD can determine the success of its UL access attempt in the previous time frame by checking MsgB of the current time frame, where access failure is identified if its ID is absent from the MsgB.

Compared to the four-step RACH procedure \cite{4-rach} commonly used in GBA, GFA's two-step RACH is subject to lower signal overhead and access latency, demonstrating superb applicability to the mMTC deployment and operation where a huge amount of miniature MTCDs. However, it increases the likelihood of collisions among uncoordinated access attempts by different MTCDs for common RBs.

\begin{figure}[btp]
    \centering
    \includegraphics[width=0.49\textwidth]{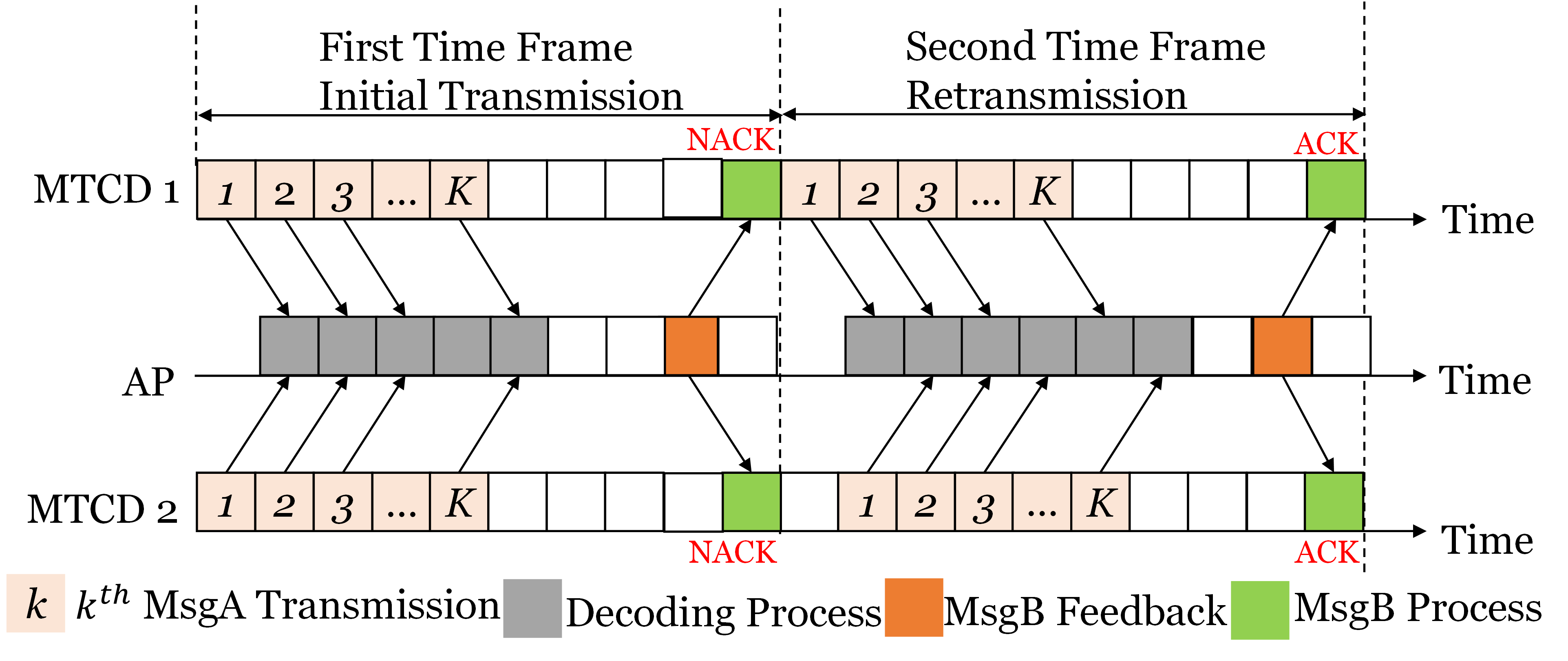}
    \caption{K-Repetition Grant-Free Transmission Procedure.}
    \label{fig: k-rep}
\end{figure}

To enhance the system robustness and throughput, 3GPP incorporates K-repetition \cite{k-repetition} with GFA, and the resultant K-GFA allows multiple copies of data as redundancy to be repeatedly transmitted in a time frame. Fig. \ref{fig: k-rep} presents the transmission procedures of two MTCDs in the conventional K-GFA scheme, where each square represents a time slot in a time frame; and each MTCD sends a packet repeatedly $K$ times to the AP which in turn independently decodes each replica. Successful delivery requires at least one of the $K$ replicas to be decoded. After processing the received replicas, MTCDs check the AP's MsgB via a broadcast channel to determine whether the data is successfully received.

\section{Proposed K-GFA Scheme}
We investigate a novel K-GFA scheme by incorporating the CRDSA-IIC mechanism with RS code, aiming at high-fidelity and low-latency K-GFA systems via robust UL random access. The section firstly introduces the media access control (MAC) protocol that supports the proposed scheme, followed by the adopted MUD mechanism based on CRDSA and IIC.

\subsection{Proposed MAC protocol} 
Recall that the traditional K-GFA system has each MTCD to transmit a packet for $K$ times within a time frame, and as long as anyone out of $K$ replicas is received, the packet is considered successfully received. The proposed CRDSA based K-GFA scheme, on the other hand, equally divides a data message consisting of $M$ packets into a number of $M/Q$ data units (DUs), each sized by $Q$ packets. By applying ($KQ$, $Q$) RS code on each DU, a RS codeword consisting of $KQ$ packets is generated for each DU and is sent within a STF that is composed of $Q$ time frames. Then in each time frame, $K$ randomly selected packets out from the $KQ$ packets of the RS codeword are launched. Carrying the packet index and the corresponding MTCD identity number (MTCD-id), each packet of the codeword is further deployed with a cyclic redundancy check (CRC) code. 

\begin{figure}[t]
    \centering
        \begin{subfigure}[b]{0.51\textwidth}
         \centering
         \includegraphics[width=\textwidth]{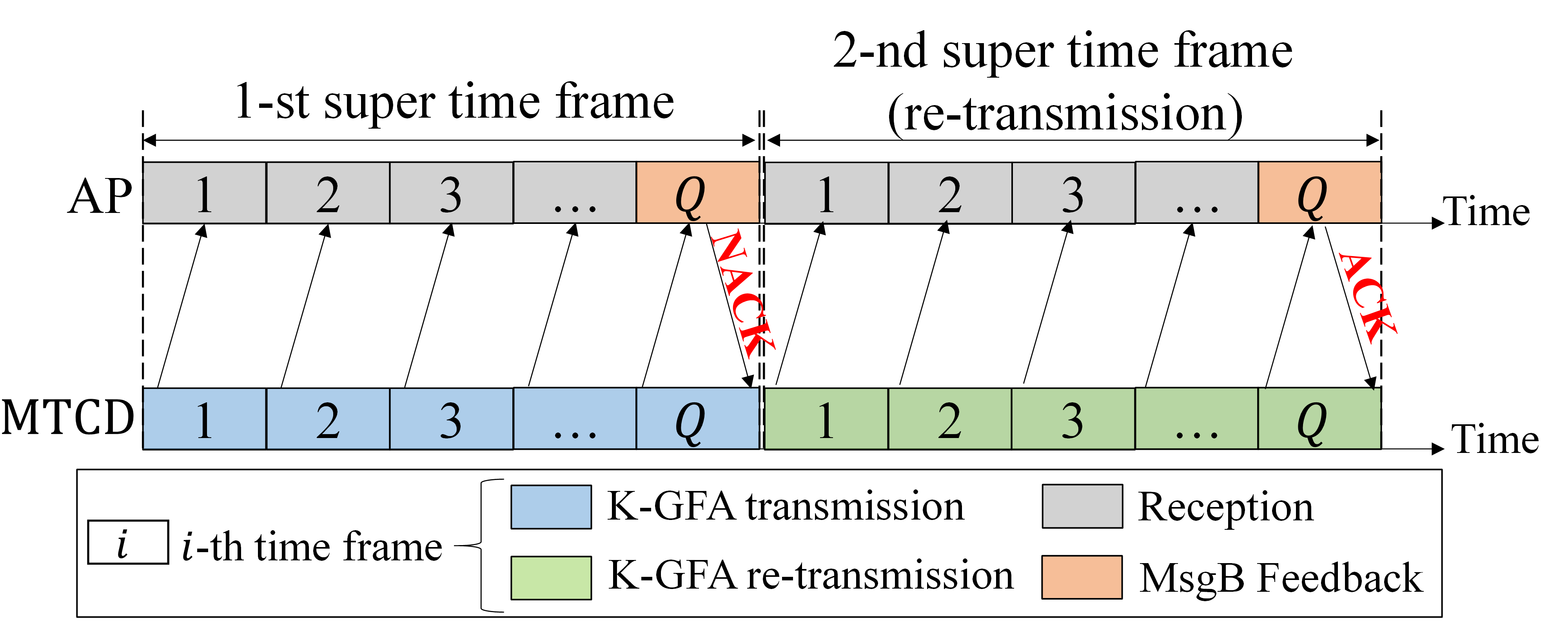}
         \caption{}
         \label{fig: STF structure}
     \end{subfigure}
    \begin{subfigure}[b]{0.49\textwidth}
         \centering
         \includegraphics[width=\textwidth]{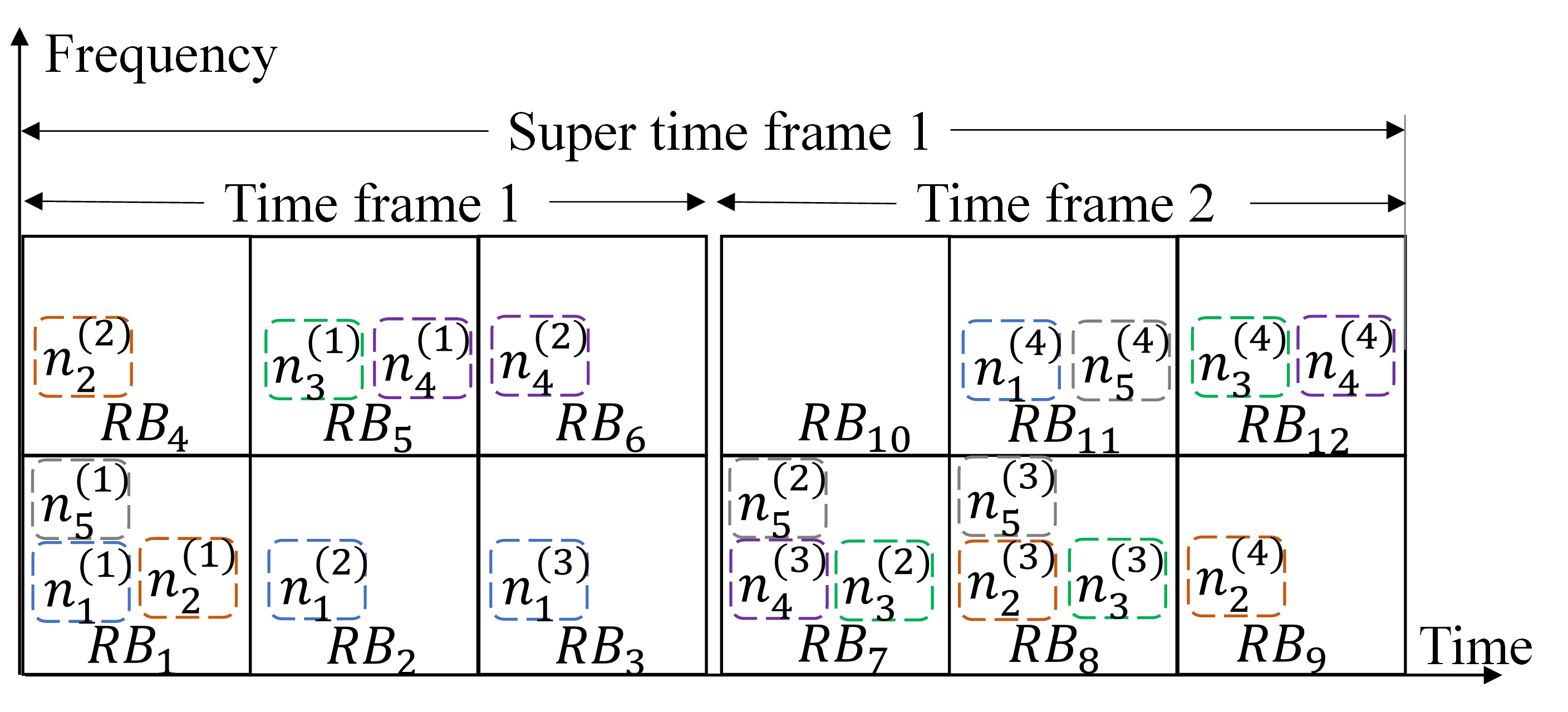}
         \caption{}
         \label{fig: resource structure}
     \end{subfigure}
     \caption{(a) Proposed K-GFA MAC structure and (b) resource structure of a super time frame and MTCDs distribution over RBs ($R=6$, $N=5$, $K=2$, $Q=2$).}
    \label{fig1}
\end{figure}

Fig. \ref{fig1}(a) shows the proposed MAC structure. The DU can be recovered by the AP if at least $Q$ out of the launched $KQ$ packets of the RS codeword in a STF are successfully obtained; otherwise re-transmission of the DU takes place in the subsequent STF in which the amount of $Q$ time frames (i.e., a STF) tops up the delay of the DU. Here, the re-transmission of each DU, triggered by NACK over the MsgB from the AP at the end of each STF, shall be taken place in the very next STF. Lastly, the data message is successfully received at the AP if all the $M/Q$ DUs are successfully recovered. 
 
The expected delay of a DU in terms of the number of time frames can be expressed as $Q/{\mathcal{P}}$, where $\mathcal{P}$ refers to the expected access probability of a DU. Thus the expected message delay denoted as $\mathcal{D}$, in the unit of the number of time frames, can be expressed as:
\begin{equation}
\mathcal{D} =  \frac{M}{Q} \frac{Q}{\mathcal{P}} = \frac{M}{\mathcal{P}}
\label{eq: delay}
\end{equation}

\subsection{CRDSA based MUD}

The following paragraphs introduce the CRDSA based MUD mechanism employed in the proposed K-GFA scheme. 

Let $N$ denote the number of MTCDs, and $n^{(p)}_i$ denote the $p$-th packet of the RS codeword of MTCD $n_i$, where $p = 1, \dots, QK$ and $i = 1, \dots, N$. An example of MTCD access map is given in Fig. \ref{fig1}, where $N=5$ MTCDs transmit $QK = 4$ packets over a STF comprised of $Q = 2$ time frames each including $R = 6$ RBs. Due to random resource selection, the MTCDs suffer from MAI at one or more RBs.

Without manipulating the power level differences among contending MTCDs, we assume only the RBs containing exclusively a single user's signal are decodable, such as $RB_2$, $RB_3$, $RB_4$ and $RB_{9}$ in Fig. \ref{fig1}(b), while the RBs with two or more MTCDs' signals cannot be decoded. We call such RBs with only a single MTCD signal as \emph{exclusive RBs}, and the corresponding packets as \emph{exclusive packets}.

The employed CRDSA based MUD scheme is deployed at the AP to potentially recover the collided RBs, where the signals of all RBs in a given STF are buffered in memory. The signals of all RBs forms a $Q \times R$ matrix denoted as $\mathbf{M}$, and the decoded signals contribute to a vector $\pmb{x}$. We assume the availability of channel status indicator (CSI) of each MTCD that is essential for an effective IC process.

The IIC process is exemplified by using Fig. \ref{fig1}, where in the STF with $R$ = 9, $N$ = 5, $K$ = 2, $Q$ = 2, the MTCD signal $n_5$, although experiencing MAI at the selected RBs, can be recovered in the proposed MUD scheme. Specifically, the first iteration yields exclusive packets $n^{(1)}_1$, $n^{(2)}_1$, $n^{(2)}_2$, $n^{(4)}_2$, by which $n_1$ and $n_2$ can be successfully obtained according to (4,2) RS code. The second iteration turns $n_5$ exclusive since two out of the four, namely $n^{(1)}_5$ and $n^{(4)}_5$ are obtained, once the former cancels the MAI composed of the replicas of $n_1$ and $n_2$ in $RB_1$, and the latter cancels the MAI composed of the replica of $n_1$ in $RB_{11}$, respectively.  

This IIC process is constrained to a maximum of $\alpha$ iterations, with the MAI signal for IC encompassing signals from no more than $\beta$ MTCDs. 

According to whether the AP is aware of which MTCD signals are contained in each RB, three IC processes are defined, namely \textit{precise IC}, \textit{context-aware IC} and \textit{blind IC}. The \textit{precise IC} can be achieved if the AP can subtract the MAI signals from their corresponding RBs. The \textit{context-aware IC} can be achieved if the AP can identify the presence of a specific MAI signal in an RB and, upon detection, selectively remove the MAI from the RB. In contrast, an AP performs \textit{blind IC} without any prior knowledge regarding which RB contains whose packet replicas. These three types of IC demonstrate different trade-offs between accuracy and complexity, making them applicable to different scenarios.

\section{Generic Implementation Model}

A generic implementation model of the proposed scheme under blind IC in each iteration, along with its complexity analysis, is given in this section.

Let $\bold{s}^{(i)}$ denote a set of the exclusive signals obtained in the $i$-th iteration, where the set size is $|\bold{s}^{(i)}|$. 
Let $\bold{c}^{(i)}$ denote the set of the signals corrected in the $i$-th iteration, where the set size is $|\bold{c}^{(i)}|$. 
Let $\bold{x}^{(i)}$ denote the set of MAI signals generated by the signals in $\bold{c}^{(i)}$, where $|\bold{x}^{(i)}|$ denotes the size of $\bold{x}^{(i)}$.
Let $\mathcal{M}^{(i)}$ denote a set of $Q \times R$ signal matrices corresponding to the result of IC out of the $(i-1)$-th iteration, where $Q$ and $R$ is the number of time frames in a STF and that of RBs in a time frame, respectively. $|\mathcal{M}^{(i)}|$ denotes the set size of $|\mathcal{M}^{(i)}|$. $\mathcal{M}^{(i)}_{j}$ refers to $j$-th matrix in $\mathcal{M}^{(i)}$, where the matrix size is $|\mathcal{M}^{(i)}_j|$.
$\mathcal{M}^{(i)}_{j, r}$ is the $r$-th element of $\mathcal{M}^{(i)}_{j}$.

\begin{figure}[ht!]
    \centering
     \centering
     \includegraphics[width=0.45\textwidth]{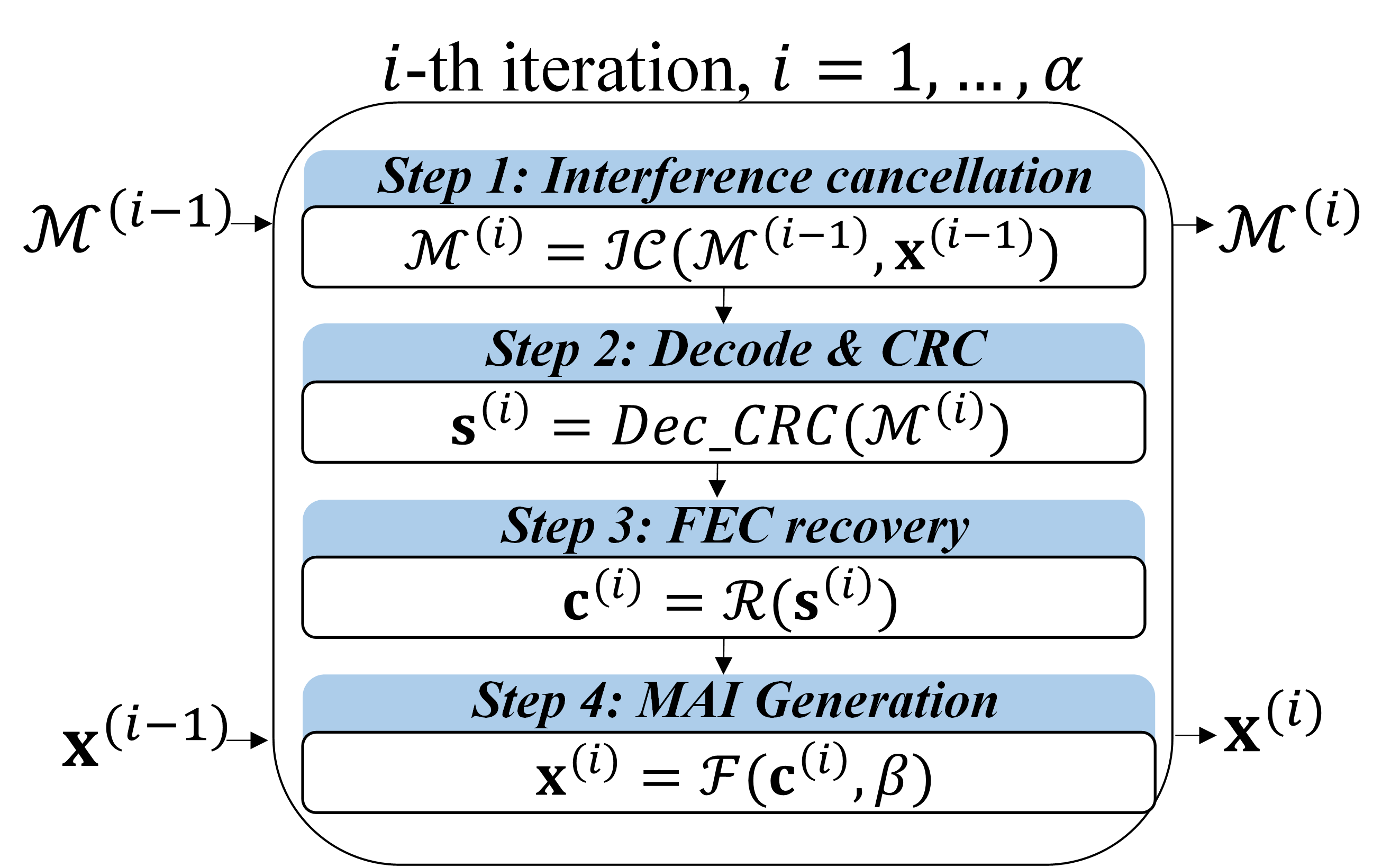}         
    \caption{A generic model for the IIC process of the proposed K-GFA scheme.}
    \label{fig2}
\end{figure}

 The four function modules for the IIC process of the proposed K-GFA scheme are (1) interference cancellation ($\mathcal{IC}$), (2) decoding-CRC (\emph{Dec\_CRC}), (3) FEC recovery ($\mathcal{R}$), and (4) MAI signal generation ($\mathcal{F}$) as shown in Fig. 3. The input of the $i$-th iteration is denoted as $\mathcal{M}^{(i-1)}$ and $\bold{x}^{(i-1)}$, producing the output $\mathcal{M}^{(i)}$ and $\bold{x}^{(i)}$, with a set of operations defined as follows.

\textit{Definition 1}: 
The IC function, denoted as $\mathcal{IC}_{t}(\mathcal{M}^{(i-1)},\bold{x}^{(i-1)}$), where $t$ can be 1, 2 and 3, corresponding to \textit{precise IC}, \textit{context-aware IC} and \textit{blind IC}, respectively, bears the following properties:
\begin{itemize}
\item The output of the function is a set of signal matrices, denoted as $\mathcal{M}^{(i)}$, where $\mathcal{M}^{(i)}_{j,r}=\mathcal{M}^{(i-1)}_{j',r}-\bold{x}^{(i-1)}_q I(\mathcal{M}^{(i-1)}_{j',r}, \bold{x}^{(i-1)}_q)$, $\forall j \in \{1\dots|\mathcal{M}^{(i)}|\}$, $\forall r \in \{1\dots QR\}$, \\ $\forall j' \in \{1\dots|\mathcal{M}^{(i-1)}|\}$, $\forall q \in \{1\dots|\bold{x}^{(i-1)}|\}$;
\item $\bold{x}^{(0)} = \left\{0\right\}$ and $\mathcal{M}^{(0)}$ contains a single raw signal matrix from RBs;
\item $|\mathcal{M}^{(i)}| \leq |\mathcal{M}^{(i-1)}| |\bold{x}^{(i-1)}|$.
\item $I(\mathcal{M}^{(i-1)}_{j',r}, \bold{x}^{(i-1)}_q)$ is a signal detection function that takes value 1 if $\mathcal{M}^{(i-1)}_{j',r}$ contains $\bold{x}^{(i-1)}_q$; value 0 otherwise. For \textit{blind IC}, it always takes value 1.
\end{itemize}

\textit{Definition 2}: The decoding function, denoted as $Dec\_CRC$($\mathcal{M}^{(i)}$), bears the properties as follows:
\begin{itemize}
\item The output of the function is a set of successfully decoded MTCD packets that have been validated by CRC, denoted as $\bold{s}^{(i)}$, where $|\bold{s}^{(i)}| \leq QKN$;
\item $\bold{s}^{(i)}_{q}$ denotes the $q$-th decoded signal of the $i$-th iteration.
\end{itemize}

\textit{Definition 3}: The RS recovery function, denoted as $\mathcal{R}$($\bold{s}^{(i)}$), bears the properties as follows:
\begin{itemize}
\item The function recovers remaining packets of MTCDs whose decoded packets involved in $\bold{s}^{(i)}$ is more than $Q$, where the output is a set denoted as $\bold{c}^{(i)}$ with the size of $|\bold{c}^{(i)}| \leq QN(K-1)$ and  $\bold{c}^{(i)} \bigcap \bold{s}^{(i)} = \emptyset$;
\item $\bold{c}^{(i)}_{p}$ denotes the $p$-th element in $\bold{c}^{(i)}$.
\end{itemize}

\textit{Definition 4}: The MAI signal generation function is denoted as $\mathcal{F}(\mathbf{c}^{(i)}, \beta)$, where the properties are as follows:
\begin{itemize}
\item The function generates a set of MAI signals, denoted as $\bold{x}^{(i)}$, by generating by superimposing up to a number of $\beta$ signals from different MTCDs in $\bold{c}^{(i)}$;
\item $\bold{x}^{(i)}_q$ denotes $q$-th MAI signal generated in the $i$-th iteration, and $\bold{x}^{(i)}_1$ is an empty signals.
\end{itemize}

Specifically, in the $i$-th iteration, IC is performed on each signal matrix in $\mathcal{M}^{(i-1)}$, denoted as $\mathcal{M}^{(i-1)}_{j}$, by subtracting each MAI signal in $\bold{x}^{(i-1)}$ from the corresponding RB. 
Each of the residual signal matrices is checked via CRC. The resulting signals are reused to retrieve collided signals of corresponding MTCDs that are retained in $\bold{c}^{(i)}$. 
Finally, a set of MAI signals, denoted as $\bold{x}^{(i)}$, is generated accordingly for IC in the subsequent iteration. 
Termination condition of the iterative process are defined as follows conditions: (1) the maximum number of iterations is reached, (2) all the MTCDs are recovered, and (3) no new MTCD is recovered and all possible aggregations of retrieved signals from the previous iteration are used.

The worst-case complexity analysis of the proposed decoding process under $/alpha=2$ and $/beta=1$, given a STF with $N$ MTCDs transmitting $KQ$ packets of an RS code, is derived in turns of the number of memory write-read, the number of IC-decoding operations, and the storage usage as follows.

\begin{itemize}
\item $1$-$st$ iteration: $N-1$ \textit{exclusive MTCDs} are retrieved from $\mathcal{M}^{(1)}_{1}$ where $Q$ out of the launched $KQ$ packets are successfully decoded for each MTCD. Then, the remaining $Q(K-1)$ packets of each MTCD are retrieved by FEC processing, contributing to the set of $Q(K-1)(N-1)$ MAI signals decoded as $\bold{x}^{(1)}$.
\item $2$-$nd$ iteration: $Q(K-1)(N-1)$ signal matrices are generated by IC with $Q(K-1)(N-1)$ MAI signals on the set $\bold{x}^{(1)}$, where the remaining MTCD can be recovered.
\end{itemize}

During the whole process, only a received signal matrix and $Q(K-1)(N-1)$ MAI signals are buffered. Thus, the storage complexity is expressed as $O(R+QKN)$. The computational complexity for memory write-read and decoding can be formulated as $O(2QKNR)$ and $O(QKNR)$, respectively.

\section{Analytical Models}
The section provides our analytical model for the proposed K-GFA scheme with $\alpha=2$ and $\beta=1$, where up to two iterations of IC and one MTCD signal taken for MAI is allowed. Given a set of $S$ events $\mathcal{A} = \left\{A_1, A_2, \dots, A_S \right\}$, the sum of probability of the difference between intersection of a number of $\mathbb{k}$ events and union of remaining is as follows
\begin{equation}
\sum\limits_{\substack{\pi \subset \mathcal{A}, \\ |\pi| = \mathbb{k}}} 
\text{P}(\bigcap\limits_{A \in \pi}A - \bigcup\limits_{\substack{A'\in \mathcal{A}/\pi}} A') = \sum^{S}_{k = \mathbb{k}}(-1)^{k-\mathbb{k}} {k \choose \mathbb{k}}
\sum\limits_{\substack{G \subset \left\{1, \dots, S\right\}, \\ |G| = k \\ \pi \subset G}} P(\bigcap_{g\in G} A_g)
\label{eq: set-difference}
\end{equation}

The involved events are defined as follows:
\begin{itemize}
\item $D_1$: MTCD $n$ is recoverable in $1$-st iteration.
\item $D_2$: MTCD $n$ originally has less than $Q$ exclusive packets and becomes recoverable after IC in $2$-nd iteration.
\end{itemize}

Given a STF with $QR$ RBs and $N$ active K-GFA MTCDs, each with $QK$ packets, the probability of $n$ for $\alpha=2$ and $\beta=1$, denoted as $\mathcal{P}(2, 1, R, N, K, Q)$ can be calculated by
\begin{equation}
\mathcal{P}(2, 1, R, N, K, Q) = P(D_1) + P(D_2)
\label{eq: analytical}
\end{equation}
where $P(D_1)$ and $P(D_2)$ are provided in \textit{Lemma 1} and \textit{Lemma 2}, respectively, that are given as follows.

\textit{Lemma 1:} Given $QR$ RBs, i.e., $r_1, \dots, r_{QR}$, the MTCD randomly selects $QK$ RBs for UL K-GFA transmission. The access probability $P(D_1)$ can be expressed as:
\begin{equation}
P(D_1) = \sum_{Q\leq \mathbb{k} \leq k \leq QK}(-1)^{k-\mathbb{k}} {k \choose \mathbb{k}} {QK \choose k} \left(\frac{{QR - k \choose QK}}{{QR \choose QK}} \right)^{N-1}
\label{eq: Lemma1}
\end{equation}
\begin{IEEEproof}
    See Appendix A
\end{IEEEproof}

\textit{Lemma 2:} Given $QR$ RBs, i.e., $r_1, \dots, r_{QR}$, the MTCD randomly selects $QK$ RBs for UL K-GFA transmission. Assuming that $R > N \gg QK$, we formulate the probability that less than $Q$ of selected is exclusive to the MTCD $n$ but becomes recoverable after IC as follows:
\begin{equation}
\begin{split}
&P(D_2) = 
\sum\limits_{\substack{
Q \leq \mathbb{k}_{n} + \mathbb{C}\\ 
1\leq \mathbb{C} \leq C \leq QK\\
0 \leq \mathbb{k}_{n} < Q \\ 
\mathbb{k}_{n} \leq k_{n} \leq QK - C\\
}}
\sum\limits_{\substack{ 
Q \leq \mathbb{k}_{1} \leq k_{1} \leq QK-1 \\ 
\dots\\
Q \leq \mathbb{k}_{C} \leq k_{C} \leq QK-1 \\ 
}} 
\frac{{QR - \mathcal{G} \choose QK}^{N - C - 1}}
{{QR \choose QK}^{N}}
{QR \choose QK + \kappa_{C}}
\\
&\mathcal{H}(N, C, \mathbb{C}, QK, \mathbb{k}_{1}, \dots, \mathbb{k}_{C}, \mathbb{k}_{n}, k_{1}, \dots, k_{C}, k_{n})
\prod^{C}_{j = 1} 
{QR - \mathcal{G} \choose QK - 1 - k_j}
\\
\label{eq: lemma 2}
\end{split}
\end{equation}
where $\kappa_{C} = \sum^{C}_{j=1}k_j$, $\mathcal{G} = k_{n} + C + \kappa_{C}$ and $\mathcal{H}(.)$ is a coefficient given by
\begin{equation}
\begin{split}
&\mathcal{H}(N, C, \mathbb{C}, QK, \mathbb{k}_{1}, \dots, \mathbb{k}_{C}, \mathbb{k}_{n}, k_{1}, \dots, k_{C}, k_{n})\\
= &(-1)^{(\substack{\mathcal{G} - \mathbb{G})}}
{N-1 \choose C} {C \choose \mathbb{C}}
{k_n \choose \mathbb{k}_n}
{QK + \kappa_{C} \choose C + \kappa_{C}}
{QK - C \choose k_n}\\
&\prod^{C}_{j = 1} (C - j + 1 + \kappa_C) {k_j \choose \mathbb{k}_j} { \sum^{C}_{q = j} k_q \choose k_j}
\end{split}
\end{equation}
where $\mathbb{K} = \sum^{C}_{j = 1} \mathbb{k}_j$ and $\mathbb{G} = \mathbb{C} + \mathbb{K} + \mathbb{k}_n$.

\begin{IEEEproof}
See Appendix B.
\end{IEEEproof}
\textit{Theorem 1 (Access probability approximation)} 

Assuming $R \geq N \gg QK$, the DU-level access probability of the proposed K-GFA system with $\alpha=2$ and $\beta=1$ can be approximated by 
\begin{equation}
\mathcal{P}(2, 1, \gamma, Q, K) = \widetilde{P}(D_1) + \widetilde{P}(D_2)
\label{eq: approximation}
\end{equation}
where $\gamma = N/R$, $\widetilde{P}(D_1)$ and $\widetilde{P}(D_2)$ refers to the approximation of the probability $P({D_1})$ and $P(D_2)$, respectively, where:
\begin{equation}
\widetilde{P}(D_1) = \sum\limits_{Q\leq \mathbb{k} \leq k \leq QK} (-1)^{k-\mathbb{k}} {k \choose \mathbb{k}} {QK \choose k} e^{-Kk\gamma}
\label{eq: approximation of D_1}
\end{equation}
and,
\begin{equation}
\begin{split}
\widetilde{P}(D_2) & = 
\sum\limits_{
\substack{
1 \leq \mathbb{C} \leq C \leq QK\\
Q \leq \mathbb{k}_n + \mathbb{C}\\
0 \leq \mathbb{k}_n < Q\\
\mathbb{k}_n \leq k_n \leq QK - C
}
}
\sum\limits_{
\substack{
Q \leq \mathbb{k}_1 \leq k_1 \leq QK - 1\\
\dots\\
Q \leq \mathbb{k}_C \leq k_C \leq QK - 1
}}
\frac{(\frac{\gamma}{Q})^{C} e^{-K\mathcal{G}\gamma}}{C!} \\
&\widetilde{\mathcal{H}}(C, \mathbb{C}, QK, \mathbb{k}_{1}, \dots, \mathbb{k}_{C}, \mathbb{k}_{n}, k_{1}, \dots, k_{C}, k_{n})
\end{split}
\label{eq: Theorem 1}
\end{equation}
where $\widetilde{\mathcal{H}}(.)$ is a coefficient given by
\begin{equation}
\begin{split}
&\widetilde{\mathcal{H}}(C, \mathbb{C}, QK, \mathbb{k}_{1}, \dots, \mathbb{k}_{C}, \mathbb{k}_{n}, k_{1}, \dots, k_{C}, k_{n}) \\
= &(-1)^{\mathcal{G} - \mathbb{G}}
{C \choose \mathbb{C}}
{k_n \choose \mathbb{k}_n}{QK + \kappa_C \choose C + \kappa_C} {QK - C \choose k_n} \frac{(QK)!}{(QK + \kappa_C)!}
\\
&\prod^{C}_{j = 1}(C - j + 1 + \kappa_C) {k_j \choose \mathbb{k}_j} {\sum^{C}_{q = j} k_q \choose k_j} \frac{(QK)!}{(QK - 1 - k_j)!}
\label{eq: approx H}
\end{split}
\end{equation}

\begin{IEEEproof}
See Appendix C. 
\end{IEEEproof}

\begin{table*}[ht!]
\centering
\caption{The accuracy of the proposed analytical model and approximation model in terms of ANA=$\frac{|P^{(analy)}-P^{(sim)}|}{P^{(sim)}}$*100\% and APP=$\frac{|P^{(approx)}-P^{(sim)}|}{P^{(sim)}}$*100\% ($P^{(sim)}$: the simulated result in term of percentage)}
\begin{tabular}{|ll|ccc|ccc|ccc|ccc|}
\hline
\rowcolor[HTML]{C0C0C0} 
\multicolumn{2}{|l|}{\cellcolor[HTML]{C0C0C0}}                               
& \multicolumn{3}{c|}{\cellcolor[HTML]{C0C0C0}0.1}                           & \multicolumn{3}{c|}{\cellcolor[HTML]{C0C0C0}0.3}                          & \multicolumn{3}{c|}{\cellcolor[HTML]{C0C0C0}0.5}                          
& \multicolumn{3}{c|}{\cellcolor[HTML]{C0C0C0}0.7}                          
\\ \cline{3-14}
\rowcolor[HTML]{C0C0C0}
\multicolumn{2}{|l|}{\multirow{-2}{*}{\cellcolor[HTML]{C0C0C0}\diagbox[width=\widthof{21}+\widthof{P1(sim)}+4\tabcolsep+\arrayrulewidth]{(Q, K)}{$\gamma$}}}       & \multicolumn{1}{c|}{\cellcolor[HTML]{C0C0C0}N=25}
& \multicolumn{1}{l|}{\cellcolor[HTML]{C0C0C0}N=100}
& N=250
& \multicolumn{1}{c|}{\cellcolor[HTML]{C0C0C0}N=25}   
& \multicolumn{1}{l|}{\cellcolor[HTML]{C0C0C0}N=100}   
& N=250
& \multicolumn{1}{c|}{\cellcolor[HTML]{C0C0C0}N=25}   
& \multicolumn{1}{c|}{\cellcolor[HTML]{C0C0C0}N=100}  
& N=250                          
& \multicolumn{1}{c|}{\cellcolor[HTML]{C0C0C0}N=25}   
& \multicolumn{1}{l|}{\cellcolor[HTML]{C0C0C0}N=100}  
& N=250                          
\\ \hline
\multicolumn{1}{|l|}{\cellcolor[HTML]{C0C0C0}}                        
& \cellcolor[HTML]{C0C0C0}$P^{(sim)}$
& \multicolumn{1}{l|}{99.9748}                                 
& \multicolumn{1}{l|}{99.9857}                                
& 99.9876                                 
& \multicolumn{1}{l|}{94.7712}                        
& \multicolumn{1}{l|}{94.343}                          
& 94.2498                        
& \multicolumn{1}{l|}{67.8784}                        
& \multicolumn{1}{l|}{66.3184}                       
& \multicolumn{1}{l|}{65.9964}   
& \multicolumn{1}{l|}{33.9908}                        
& \multicolumn{1}{l|}{34.4265}                       
& 34.7588                        
\\ \cline{2-14} 
\rowcolor[HTML]{FFFFFF} 
\multicolumn{1}{|l|}{\cellcolor[HTML]{C0C0C0}}                        
& \cellcolor[HTML]{C0C0C0}ANA                          
& \multicolumn{1}{c|}{\cellcolor[HTML]{FFFFFF}0.0219}          
& \multicolumn{1}{l|}{\cellcolor[HTML]{FFFFFF}0.0047}          
& 0.0012                                  
& \multicolumn{1}{c|}{\cellcolor[HTML]{FFFFFF}0.1816} 
& \multicolumn{1}{l|}{\cellcolor[HTML]{FFFFFF}0.0428} 
& 0.0182                         
& \multicolumn{1}{c|}{\cellcolor[HTML]{FFFFFF}0.2469} 
& \multicolumn{1}{c|}{\cellcolor[HTML]{FFFFFF}0.0336} 
& 0.0165                         
& \multicolumn{1}{c|}{\cellcolor[HTML]{FFFFFF}0.2609} 
& \multicolumn{1}{l|}{\cellcolor[HTML]{FFFFFF}0.0196} 
& 0.0099                         
\\ \cline{2-14} 
\rowcolor[HTML]{FFFFFF} 
\multicolumn{1}{|l|}{\multirow{-3}{*}{\cellcolor[HTML]{C0C0C0}(2,2)}} 
& \cellcolor[HTML]{C0C0C0}APP.                         
& \multicolumn{1}{c|}{\cellcolor[HTML]{FFFFFF}0.0127}          
& \multicolumn{1}{l|}{\cellcolor[HTML]{FFFFFF}0.0019}          
& \cellcolor[HTML]{FFFFFF}\textless{}1e-4 
& \multicolumn{1}{c|}{\cellcolor[HTML]{FFFFFF}0.7226} 
& \multicolumn{1}{l|}{\cellcolor[HTML]{FFFFFF}0.1907}  
& \cellcolor[HTML]{FFFFFF}0.0758 
& \multicolumn{1}{c|}{\cellcolor[HTML]{FFFFFF}3.0818} 
& \multicolumn{1}{c|}{\cellcolor[HTML]{FFFFFF}0.8020} 
& \cellcolor[HTML]{FFFFFF}0.3180 
& \multicolumn{1}{c|}{\cellcolor[HTML]{FFFFFF}3.1124} 
& \multicolumn{1}{l|}{\cellcolor[HTML]{FFFFFF}0.7595} 
& \cellcolor[HTML]{FFFFFF}0.2900 
\\ \hline
\multicolumn{1}{|l|}{\cellcolor[HTML]{C0C0C0}}                        
& \cellcolor[HTML]{C0C0C0}$P^{(sim)}$
& \multicolumn{1}{c|}{1}                                       
& \multicolumn{1}{c|}{1}                                       
& 1                                       
& \multicolumn{1}{l|}{98.2152}                       
& \multicolumn{1}{l|}{97.9717}                         
& 97.885                         
& \multicolumn{1}{l|}{65.5196}                        
& \multicolumn{1}{l|}{64.2014}                        
& \multicolumn{1}{l|}{63.8352}   
& \multicolumn{1}{l|}{20.5404}                        
& \multicolumn{1}{l|}{22.2352}                        
& 22.8848                        
\\ \cline{2-14} 
\rowcolor[HTML]{FFFFFF} 
\multicolumn{1}{|l|}{\cellcolor[HTML]{C0C0C0}}                        
& \cellcolor[HTML]{C0C0C0}ANA                          
& \multicolumn{1}{c|}{\cellcolor[HTML]{FFFFFF}0.0003}          
& \multicolumn{1}{l|}{\cellcolor[HTML]{FFFFFF}0.0001}          
& \cellcolor[HTML]{FFFFFF}\textless{}1e-4 
& \multicolumn{1}{c|}{\cellcolor[HTML]{FFFFFF}0.3077} 
& \multicolumn{1}{l|}{\cellcolor[HTML]{FFFFFF}0.0645} 
& \cellcolor[HTML]{FFFFFF}0.0484 
& \multicolumn{1}{c|}{\cellcolor[HTML]{FFFFFF}0.5813} 
& \multicolumn{1}{c|}{\cellcolor[HTML]{FFFFFF}0.0177} 
& \cellcolor[HTML]{FFFFFF}0.0387 
& \multicolumn{1}{c|}{\cellcolor[HTML]{FFFFFF}0.2120} 
& \multicolumn{1}{l|}{\cellcolor[HTML]{FFFFFF}0.0730} 
& \cellcolor[HTML]{FFFFFF}0.1772 \\ \cline{2-14} 
\rowcolor[HTML]{FFFFFF} 
\multicolumn{1}{|l|}{\multirow{-3}{*}{\cellcolor[HTML]{C0C0C0}(2,3)}} 
& \cellcolor[HTML]{C0C0C0}APP.                         
& \multicolumn{1}{c|}{\cellcolor[HTML]{FFFFFF}\textless{}1e-4} 
& \multicolumn{1}{l|}{\cellcolor[HTML]{FFFFFF}\textless{}1e-4} 
& \cellcolor[HTML]{FFFFFF}\textless{}1e-4 
& \multicolumn{1}{c|}{\cellcolor[HTML]{FFFFFF}0.4253} 
& \multicolumn{1}{l|}{\cellcolor[HTML]{FFFFFF}0.1275}  
& \cellcolor[HTML]{FFFFFF}0.0291 
& \multicolumn{1}{c|}{\cellcolor[HTML]{FFFFFF}2.8657} 
& \multicolumn{1}{c|}{\cellcolor[HTML]{FFFFFF}0.8713} 
& \cellcolor[HTML]{FFFFFF}0.3027 
& \multicolumn{1}{c|}{\cellcolor[HTML]{FFFFFF}2.5565} 
& \multicolumn{1}{l|}{\cellcolor[HTML]{FFFFFF}0.5471} 
& \cellcolor[HTML]{FFFFFF}0.0119 
\\ \hline
\multicolumn{1}{|l|}{\cellcolor[HTML]{C0C0C0}}                        
& \cellcolor[HTML]{C0C0C0}$P^{(sim)}$
& \multicolumn{1}{l|}{99.9992}                                 
& \multicolumn{1}{l|}{99.9988}                                 
& 99.9994                                 
& \multicolumn{1}{l|}{96.3084}                        
& \multicolumn{1}{l|}{96.002}                          
& 95.9084                        
& \multicolumn{1}{l|}{61.594}                         
& \multicolumn{1}{l|}{59.5384}                        
& \multicolumn{1}{l|}{59.0484}   
& \multicolumn{1}{l|}{21.3376}                        
& \multicolumn{1}{l|}{21.8367}                        
& 22.1744                        
\\ \cline{2-14} 
\rowcolor[HTML]{FFFFFF} 
\multicolumn{1}{|l|}{\cellcolor[HTML]{C0C0C0}}                        
& \cellcolor[HTML]{C0C0C0}ANA                          
& \multicolumn{1}{c|}{\cellcolor[HTML]{FFFFFF}0.0014}          
& \multicolumn{1}{l|}{\cellcolor[HTML]{FFFFFF}0.0013}          
& 0.0004                                  
& \multicolumn{1}{c|}{\cellcolor[HTML]{FFFFFF}0.3683} 
& \multicolumn{1}{l|}{\cellcolor[HTML]{FFFFFF}0.0997}  
& 0.0576                         
& \multicolumn{1}{c|}{\cellcolor[HTML]{FFFFFF}0.0719} 
& \multicolumn{1}{c|}{\cellcolor[HTML]{FFFFFF}0.1765} 
& 0.0810                         
& \multicolumn{1}{c|}{\cellcolor[HTML]{FFFFFF}1.1299} 
& \multicolumn{1}{l|}{\cellcolor[HTML]{FFFFFF}0.0359} 
& 0.0751                         
\\ \cline{2-14} 
\rowcolor[HTML]{FFFFFF} 
\multicolumn{1}{|l|}{\multirow{-3}{*}{\cellcolor[HTML]{C0C0C0}(3,2)}} 
& \cellcolor[HTML]{C0C0C0}APP.                        
& \multicolumn{1}{c|}{\cellcolor[HTML]{FFFFFF}0.0003}          
& \multicolumn{1}{l|}{\cellcolor[HTML]{FFFFFF}0.0006}          
& \cellcolor[HTML]{FFFFFF}\textless{}1e-4 
& \multicolumn{1}{c|}{\cellcolor[HTML]{FFFFFF}0.5495} 
& \multicolumn{1}{l|}{\cellcolor[HTML]{FFFFFF}0.1392}  
& 0.0387                         
& \multicolumn{1}{c|}{\cellcolor[HTML]{FFFFFF}4.4982} 
& \multicolumn{1}{c|}{\cellcolor[HTML]{FFFFFF}0.9779} 
& 0.3811                         
& \multicolumn{1}{c|}{\cellcolor[HTML]{FFFFFF}4.0291} 
& \multicolumn{1}{l|}{\cellcolor[HTML]{FFFFFF}0.7563} 
& 0.2504                         
\\ \hline
\end{tabular}
\end{table*}

\section{Case Study}

Extensive numerical experiments are conducted to validate both the proposed analytical model on access probability for the case of $\alpha=2$ and $\beta=1$, as well as to investigate the impact by various key parameters to the performance in terms of access probability of each DU and message delay, including, $K$, $\gamma$, and $Q$. The message delay is defined as the expected number of time frames required for successful transmission of a message with a size $M$ packets. 

Table I shows the normalized difference between the derived access probabilities and the corresponding random simulation results. We obtain the following observations. Firstly, the proposed analytical model achieves very close access probability performance to that by the simulation (within $0.1\%$ of deviation from the simulation result), particularly when $N$ is large and $\gamma$ is small. Secondly, using $\gamma=N/R$ as the metric instead of individual values of $N$ and $R$ can effectively characterize the access probability performance, which is given by the approximate model (generally within $1\%$ of deviation from the simulation result).

We examine the access probability for each DU (i.e., DU-level delay) that contains a message of 32 packets. The message-level delay is defined as the latency for successfully receiving all the 32 packets of the original message, and in the ideal case such message-level delay is 32 time frames. The message-level delay can be calculated using \eqref{eq: delay}, where $\mathcal{P}$ is given by \eqref{eq: analytical}.

\begin{figure}[ht!]
    \centering
    \includegraphics[width=0.49\textwidth]{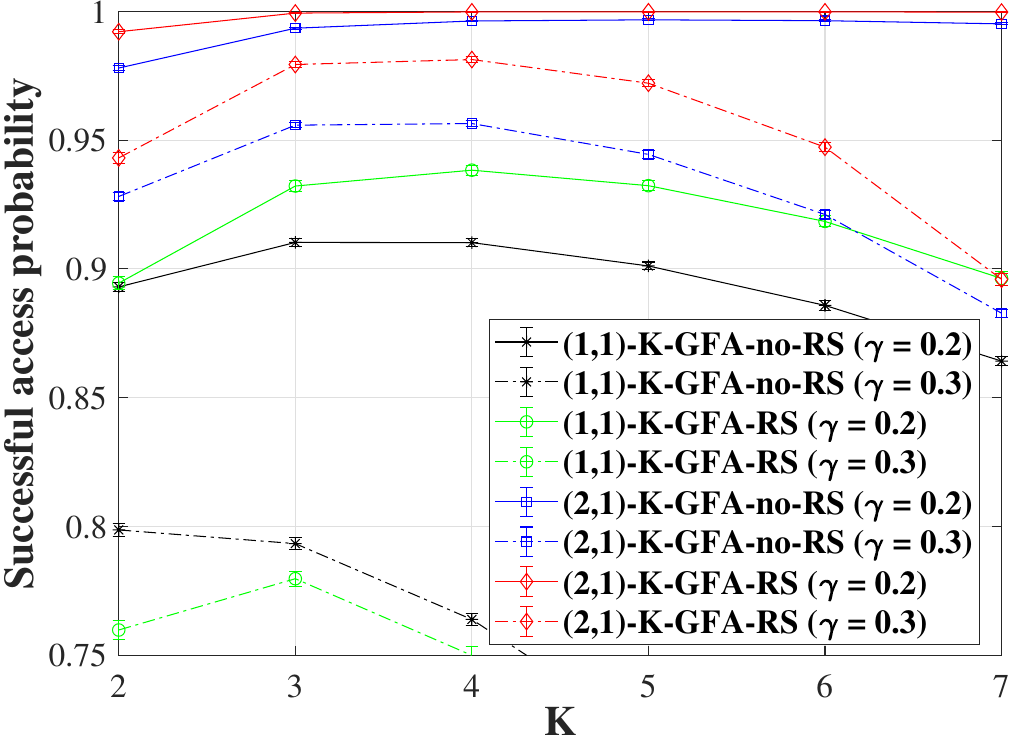}
    \caption{Excepted DU-level access probability of a MTCD for different K-GFA systems versus various values of $K$ under $N$ = 100 with $\gamma = 0.2$ and 0.3}
    \label{fig: result_1}
\end{figure}
\begin{figure}[ht!]
    \centering
    \begin{subfigure}[b]{0.47\textwidth}
         \centering
         \includegraphics[width=\textwidth]{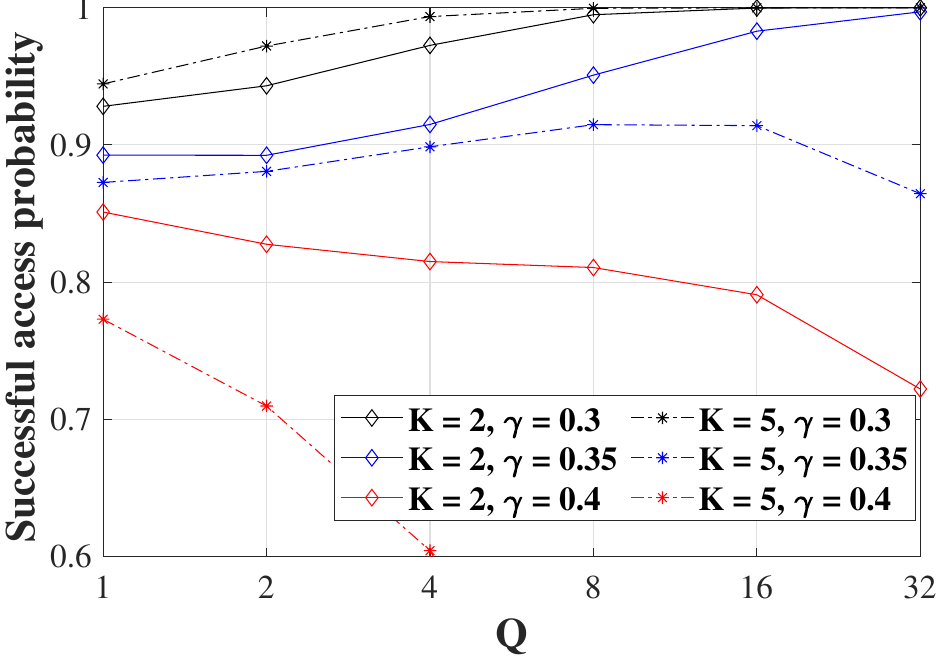}
         \label{fig: sap}
         \caption{}
     \end{subfigure}
     \begin{subfigure}[b]{0.47\textwidth}
         \centering
         \includegraphics[width=\textwidth]{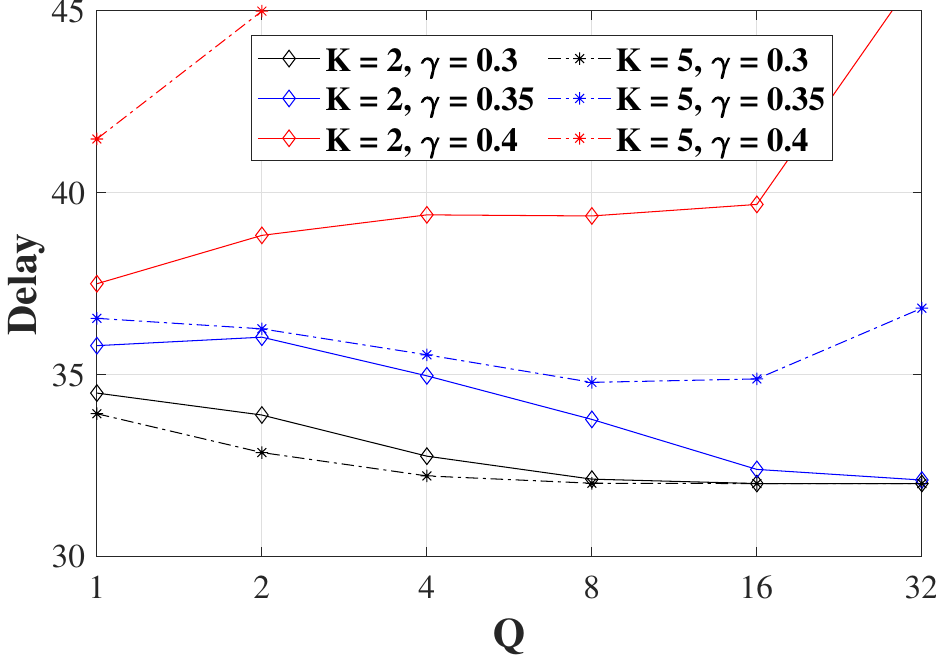}
         \label{fig: delay}
         \caption{}
     \end{subfigure}
    \caption{Performance of (2,1)-K-GFA-RS system in terms of (a) excepted DU-level access probability and (b) message  delay versus various values of $Q$ under $N$ = 100, $K = 2$, 5 with $\gamma = $ 0.3, 0.35 and 0.4}
    \label{fig: result_2}
\end{figure}

Fig. \ref{fig: result_1} shows the performance of the proposed scheme denoted as (2,1)-K-GFA-RS and is compared to three counterparts, namely (1,1)-K-GFA-no-RS (without IIC or RS code), (1,1)-K-GFA-RS (without IIC while with RS code) and (2,1)-K-GFA-no-RS (with IIC while without RS code), respectively. By taking $Q=2$ and increasing $K$ from 1 to 7, we see that (2,1)-K-GFA-RS outperforms the other three under various $\gamma$, clearly indicating the use of RS code and the CDRSA based IIC mechanism can solidly contribute to the DU-level access probability performance. We have also observed there exists a value of $K$ in each case for achieving the optimal performance. For example, with (2,1)-K-GFA-RS we should take $K=5$ when $\gamma=0.2$, and $K=4$ when $\gamma=0.3$.

Figs. \ref{fig: result_2}(a) and (b) demonstrate the DU-level access probability and the message-level delay of (2,1)-K-GFA-RS, respectively, by varying $Q$ values from 1 to 32, under different $\gamma$ and $K$ values. Firstly as shown in Fig. \ref{fig: result_2}(a), increasing $Q$ results in an improvement on the DU-level access probability for smaller $\gamma$; while such an effect is reversed when $\gamma$ exceeds a certain point (e.g., 0.4). This is due to the challenge of securing $Q$ \textit{exclusive RBs} for a MTCD initially when dealing with a large $\gamma$. Then, Fig. \ref{fig: result_2}(b) shows a similar trend from the perspective of message-level delay, which is inversely proportional to the DU-level access probability as in \eqref{eq: delay} disregard the value of $\gamma$. The access probability becomes decreased when $\gamma$ is large as indicated in Fig. \ref{fig: result_2}(a), leading to the fact that the larger $Q$ the longer the message-level delay. We also find that the user intensity $\gamma$ affects the selection of operation parameters. For example, when $K = 5$, the optimal $Q$ for given $\gamma = 0.3, 0.35, 0.4$ are 32, 8, 1, respectively. From the operational perspective, the AP can determine the optimal values of $Q$ and $K$ according to observed $\gamma$ and notify the MTCDs via MsgB for achieving the optimal message-level delay and DU-level access probability performance.

\section{Conclusions}

The paper has introduced a novel K-GFA scheme by incorporating the iterative interference cancellation (IIC) mechanism of contention resolution diversity Slotted Aloha (CRDSA) with Reed-Solomon (RS) code, for achieving effective multi-user detection (MUD) in the presence of uncoordinated access by miniature mMTC devices (MTCDs). Our contributions are in several folds. Firstly, we defined a transmission structure of MAC protocol for $K$ replicas of each data message that can accommodate the RS code deployment. Secondly, we came up with a generic implementation model for the blind IC scenario. Thirdly, we provided an analytical model as well as an approximate model, proving that the system can be described in terms of $\gamma=N/R$ rather than individually $N$ and $R$. Extensive numerical experiment results validated the proposed analytical and approximate models, and provided insights on the performance of the proposed K-GFA system in terms of access probability and message delay by manipulating various key parameters such as $Q$, $K$, and $\gamma$. 

\section*{appendix A}
\emph{Proof (Lemma 1)}: let $\pmb{R}$ be a given set of $QR$ RBs. The event $D_1$ is equal to the event that arbitrarily $Q$ or more selected RBs are exclusive to the target MTCD $n$ before IC. Thus, the probability of $D_1$ can be expressed as
\begin{equation}
P(D_1) = \sum_{Q\leq \mathbb{k} \leq QK}P(X_{n}(\mathbb{k}))
\label{eq: lemma1-step0}
\end{equation}
where $X_{n}(\mathbb{k})$ is event that the MTCD $n$ exactly has a number of $\mathbb{k}$ exclusive RBs, whose probability can be expressed as follows:
\begin{equation}
P(X_{n}(\mathbb{k})) = 
\sum\limits_{\substack{
\pi \subset \pmb{R} \\
|\pi| = \mathbb{k}
}}
P(\bigcap_{r \in \pi} E_{n}(r) - \bigcup_{r' \in \pmb{R}/\pi} E_{n}(r'))
\label{eq: lemma1-step1}
\end{equation}
where $E_n(r)$ refers to the event that the RB $r$ is exclusive to MTCD $n$. By using \eqref{eq: set-difference}, the \eqref{eq: lemma1-step1} is turned into as follows:
\begin{equation}
P(X_{n}(\mathbb{k})) = \sum_{\mathbb{k} \leq k \leq QK}(-1)^{(k - \mathbb{k})}
{k \choose \mathbb{k}} 
\sum\limits_{\substack{I \subset \pmb{R}, \\ |I| = k }} P(\bigcap_{r \in I} E_n(r))
\label{eq: lemma1-step2}
\end{equation}
The intersection of events $E_n(r)$ with $ r\in I$ can be turned into a new event that the MTCD $n$ is exclusive in the tagged RBs corresponding to $I$. Its probability can be expressed as follows:
\begin{equation}
P(\bigcap_{i \in I} E_n(r_i)) = 
\frac{{QR - k \choose QK-k}{QR - k \choose QK}^{(N-1)}}{{QR \choose QK}^N}, \forall I \subset \left\{1, \dots, QR\right\}, |I| = k
\label{eq: lemma1-step3}
\end{equation}
Substituting \eqref{eq: lemma1-step3} and \eqref{eq: lemma1-step2} into \eqref{eq: lemma1-step0}, we can obtain a new equation as follows:
\begin{equation}
\begin{split}
P(D_1) & = \sum_{Q \leq \mathbb{k} \leq k \leq QK} (-1)^{(k - \mathbb{k})}
{k \choose \mathbb{k}} {QR \choose k} \frac{{QR - k \choose QK-k}{QR - k \choose QK}^{(N-1)}}{{QR \choose QK}^N}
\end{split}
\end{equation}
which can be turned into \eqref{eq: Lemma1} in \emph{Lemma 1}. Q.E.D.
\section*{Appendix B}
\textit{Proof (Lemma 2)}: let $\pmb{U}$ be the set of $N-1$ active MTCDs excluding the MTCD $n$. Let $\sim D_1$ be the event that the MTCD $n$ has less than $Q$ \emph{exclusive RBs} before IC, which is equal to the union of events $X_n(\mathbb{k}_n)$ for $\mathbb{k}_n < Q$. Let $Y_n(\mathbb{C})$ be the event that the MTCD $n$ has exactly $\mathbb{C}$ \emph{retrieved RBs} with each one corresponding to a different recoverable MTCD, and $F_n$ be the union of events $Y_n(\mathbb{C})$ for $\mathbb{C} > 0$. Let $\mathcal{Z}$ be the event that the sum of \emph{exclusive RBs} and \emph{retrieved RBs} of the MTCD $n$ equals to or is more than $Q$.

We first formulate the probability of $Y_n(\mathbb{C})$ with the condition that MTCD $n$ has $k_n$ exclusive RBs as follows:
\begin{equation}
P(Y_n(\mathbb{C}) | k_n) = 
\sum\limits_{\substack{
J \subset \pmb{U} \\
|J| = \mathbb{C}
}}
P(\bigcap_{j \in J} \mathcal{T}_{n}(j) - \bigcup_{j' \in \pmb{U}/J} \mathcal{T}_{n}(j') | k_n)
\label{eq: lemma2-step0}
\end{equation}
where $\mathcal{T}_{n}(j)$ refers to the event that the MTCD $n$ is coupled with a recoverable MTCD $j$. Based on \eqref{eq: set-difference}, \eqref{eq: lemma2-step0} is turned into as follows:
\begin{equation}
\begin{split}
P(Y_n(\mathbb{C}) | k_n) &= \sum_{\mathbb{C} \leq C \leq QK}(-1)^{(C - \mathbb{C})}
{C \choose \mathbb{C}} 
{N - 1 \choose C}
P(\bigcap^{C}_{j = 1} \mathcal{T}_{n}(j) | k_n)\\
\end{split}
\label{eq: lemma2-step1}
\end{equation}
where the probability $P(\bigcap^{C}_{j = 1} \mathcal{T}_{n}(j))$ can be expressed as follows:
\begin{equation}
\begin{split}
&P(\bigcap^{C}_{j = 1} \mathcal{T}_{n}(j) | k_n) = 
\sum\limits_{
\substack{
Q \leq \mathbb{k}_1 \leq QK - 1\\ 
\dots\\
Q \leq \mathbb{k}_C \leq QK - 1
}}
P(X_1(\mathbb{k}_1), \dots, X_C(\mathbb{k}_C)| k_n)\\
&=
\sum\limits_{
\substack{
Q \leq \mathbb{k}_1 \leq k_1 \leq QK - 1\\ 
\dots\\
Q \leq \mathbb{k}_C \leq k_C \leq QK - 1
}} (-1)^{(\kappa_{C} - \mathbb{K})}
\frac{{QR - \mathcal{G} \choose QK}^{N - C - 1} }
{ {QR - k_n \choose QK}^{(N-1)}}
\frac{{QR - k_n \choose \kappa_C + C} {QR - \mathcal{G}\choose QK - C - k_n}}
{ {QR - k_n \choose QK - k_n} } 
\\
&
\prod^{C}_{j = 1} 
(C + \kappa_C - j + 1) 
{ k_j \choose \mathbb{k}_j} 
{\sum^{C}_{q = j} k_q \choose k_j} 
{QR - \mathcal{G} \choose QK - 1 - k_j}
\end{split}
\label{eq: lemma2-step2}
\end{equation}
where $\kappa_C = \sum^{C}_{j = 1}k_j$, $\mathbb{K} = \sum^{C}_{j = 1}\mathbb{k}_j$, and $\mathcal{G} = C + \kappa_{C} + k_n$

Under the assumption that $R > N \gg QK$, the event $D_2$ can be simplified as the intersection of three events: 1) $\sim D_1$; 2) $F_n$ and 3) $\mathcal{Z}$. Thus, the probability of $D_2$ can be expressed as follows:
\begin{equation}
\begin{split}
& P (D_2)= P(F_n, \sim D_1, Z)\\
& = \sum\limits_{ \substack{
Q \leq \mathbb{k}_n + \mathbb{C}\\
0 < \mathbb{C} \leq QK \\
0 \leq \mathbb{k}_n < Q
}} P(Y_n(\mathbb{C}), X_n(\mathbb{k}_n))\\
& = \sum\limits_{ \substack{
Q \leq \mathbb{k}_n + \mathbb{C}\\
0 < \mathbb{C} \leq QK \\
0 \leq \mathbb{k}_n < Q\\
\mathbb{k}_n\leq k_n \leq QK-C
}}
(-1)^{(k_n - \mathbb{k}_n)} P(Y_n(\mathbb{C})|k_n){k_n \choose \mathbb{k}_n} \\
&\frac{{QR \choose k_n}{QR - k_n \choose QK - k_n}{QR - k_n \choose QK}^{(N-1)}}
{{QR \choose QK}^{N}}
\label{eq: lemma2-step3}
\end{split}
\end{equation}
Substituting \eqref{eq: lemma2-step1} and \eqref{eq: lemma2-step2} into
\eqref{eq: lemma2-step3}, we can obtain \eqref{eq: lemma 2} in \emph{Lemma 2}. Q.E.D.
\section*{Appendix C}

\textit{Proof (Theorem 1)}: The probability mass function (PMF) of a binomial distribution can be approximated by the PMF of a Poisson distribution:
\begin{equation}
{n \choose k} p^k q^{n-k} \approx \frac{\lambda^{k} e^{-\lambda}}{k!}
\end{equation}
where $n\gg k$, $\lambda = \frac{k}{n}$, $p = \frac{k}{n^2}$, and $q = 1 - p$. Consider the limit: 
\begin{equation}
\left\{
\begin{split}
&\lim_{n\to\infty} q = 1\\
&\lim_{n\to\infty} \lambda = 0
\end{split}
\right.
\end{equation}
the binomial coefficient ${n \choose k}$ can be approximated as follows: 
\begin{equation}
{n \choose k} \approx \frac{n^{k}}{k!}
\end{equation}
Based on this, the quotient of two binomial coefficients ${n_1 \choose k_1}$ and ${n_2 \choose k_2}$ can be expressed as
\begin{equation}
\frac{{n_1 \choose k_1}}{{n_2 \choose k_2}} = \frac{n_1^{k_1}}{n_2^{k_2}}\frac{k_2!}{k_1!}
\end{equation}
Thus the probability $P(D_1)$ can be approximated as follows:
\begin{equation}
\begin{split}
\widetilde{P}(D_1) = \sum_{Q\leq \mathbb{k} \leq k \leq QK}(-1)^{k-\mathbb{k}} {k \choose \mathbb{k}} {QK \choose k} \left( 1- \frac{{k}}{{QR}} \right)^{QKN}
\end{split}
\end{equation}
where $\widetilde{P}(D_1)$ denotes the approximation of $P({D_1})$. Consider the binomial approximation: 
\begin{equation}
(1+{x})^a = e^{ax}
\end{equation}
where $|x|$ is small and $|a|$ is large. Let $x = -\frac{k}{QR}$ and $a = QKN$, the $\widetilde{P}(D_1)$ can be further expressed as \eqref{eq: approximation of D_1} in \textit{Theorem 1}.

Similarly, based on the assumption that $R \geq N \gg QK$, the binomial coefficient related to $R$ in $P(D_2)$ can be approximated as follows:
\begin{equation}
\left\{
\begin{aligned}
&\frac{{QR - \mathcal{G} \choose QK}}{{QR \choose QK}} \approx (1 - \frac{\mathcal{G}}{QR})^{QK}\\
&\frac{{QR \choose QK + \kappa_{C}}}{{QR \choose QK}} 
\approx (QR)^{\kappa_{C}} \frac{(QK)!}{(QK + \kappa_{C})!}
\\
&\prod^{C}_{j = 1}\frac{{RR - \mathcal{G} \choose QK - \Delta_j}}{{QR \choose QK}} \approx
\left(1 - \frac{\mathcal{G}}{QR}\right)^{QKC - C - \kappa_{C}} 
\left(\frac{1}{QR}\right)^{C + \kappa_{C}}
\prod^{C}_{j = 1}\frac{(QK)!}{(QK - \Delta_{n,j})!} 
\end{aligned}
\right.
\end{equation}
where $\Delta_{j} = 1 + k_j$. Thus, the approximation of $P(D_2)$ can be expressed as follows:
\begin{multline}
\widetilde{P}(D_2) = 
\sum\limits_{
\substack{
1 \leq \mathbb{C} \leq C \leq QK\\
Q \leq \mathbb{k}_n + \mathbb{C}\\
0 \leq \mathbb{k}_n < Q\\
\mathbb{k}_n \leq k_n \leq QK - C
}
}
\sum\limits_{
\substack{
Q \leq \mathbb{k}_1 \leq k_1 \leq QK - 1\\
\dots\\
Q \leq \mathbb{k}_C \leq k_C \leq QK - 1
}}
\left(1 - \frac{ \mathcal{G} }{QR}\right)^{(QK-1)N - QK -\kappa_{C}}\\
\underbrace{
\left(1 - \frac{\mathcal{G} }{QR}\right)^{N-{C}} \left(\frac{1}{QR}\right)^{{C}} {N-1 \choose C} }_{T_{1}} \\
\widetilde{\mathcal{H}}(C, \mathbb{C}, QK, \mathbb{k}_{1}, \dots, \mathbb{k}_{C}, \mathbb{k}_{n}, k_{1}, \dots, k_{C}, k_{n}) 
\label{eq: approxi 2}
\end{multline}
where $\widetilde{\mathcal{H}}$($\dots$) can be calculated by \eqref{eq: approx H}. Due to $N \gg C$, the term $T_1$ in \eqref{eq: approxi 2} can be approximated as follows:
\begin{equation}
\begin{split}
T_{1} &
= \frac{{N-1 \choose C}}{{N \choose {C}}}{N \choose {C}}\left(\frac{1}{\mathcal{G}}\right)^{{C}}
\left(\frac{\mathcal{G}}{QR}\right)^{{C}}\left(1-\frac{\mathcal{G}}{QR}\right)^{N - {C}}\\
&\approx
\frac{(N-1)^{C}}{N^{C}}\left(\frac{1}{\mathcal{G}}\right)^{{C}} 
\frac{(\frac{\mathcal{G}\gamma}{Q})^{{C}} 
e^{-\frac{\mathcal{G}\gamma}{Q}}}
{{C}!}\\
&\approx \frac{(\frac{\gamma}{Q})^{{C}}e^{-\frac{\mathcal{G}\gamma}{Q}}}{C!} \quad (\because \lim_{N\to\infty} \frac{(N-1)^{C}}{N^{{C}}} = 1)
\label{eq: T1}
\end{split}
\end{equation}
Substituting \eqref{eq: T1} into \eqref{eq: approxi 2}, we can obtain a new equation:
\begin{equation}
\begin{split}
&\widetilde{P}(D_2) = 
\sum\limits_{
\substack{
1 \leq \mathbb{C} \leq C \leq QK\\
Q \leq \mathbb{k}_n + \mathbb{C}\\
0 \leq \mathbb{k}_n < Q\\
\mathbb{k}_n \leq k_n \leq QK - C
}
}
\sum\limits_{
\substack{
Q \leq \mathbb{k}_1 \leq k_1 \leq QK - 1\\
\dots\\
Q \leq \mathbb{k}_C \leq k_C \leq QK - 1
}}
\frac{(\frac{\gamma}{Q})^{C}
e^{-\frac{\mathcal{G}((QK-1)N - QK -\kappa_{C})}{QR} - \mathcal{G}\frac{\gamma}{Q}}}{C!}
\\
&\widetilde{\mathcal{H}}(C, \mathbb{C}, QK, \mathbb{k}_{1}, \dots, \mathbb{k}_{C}, \mathbb{k}_{n}, k_{1}, \dots, k_{C}, k_{n})
 \end{split}
 \label{eq: approxi 3}
\end{equation}
Consider the approximation that
\begin{equation}
\begin{split}
\frac{\mathcal{G}((QK-1)N - QK -\kappa_{C})}{QR} \approx 
{\mathcal{G}(QK-1)\frac{\gamma}{Q}}
\end{split}
\end{equation}
\eqref{eq: approxi 3} can be turned into the \eqref{eq: Theorem 1} in \textit{Theorem 1}.
Q.E.D
\bibliographystyle{IEEEtran}
\bibliography{main.bib}
\end{document}